\documentclass[onecolumn, numberedappendix]{aastex63}
\usepackage{graphicx}
\usepackage{epsfig}
\usepackage{natbib}
\usepackage{amssymb}
\usepackage{amsbsy}
\usepackage{natbib}
\usepackage{url}
\usepackage[mathcal]{euscript}
\usepackage{color}

\newcommand{\msun}{M$_{\sun}$}

\newcommand{\etal}{et~al.}

\begin{document}
 

\title{Even More Rapidly Rotating Pre-Main Sequence M Dwarfs with
Highly Structured Light Curves: An Initial Survey  in the Lower
Centaurus-Crux and Upper Centaurus-Lupus Associations}
\author[0000-0003-3595-7382]{John Stauffer}
\affiliation{Spitzer Science Center (SSC), IPAC, California Institute of
Technology, Pasadena, CA 91125, USA}
\author[0000-0001-6381-515X]{Luisa M. Rebull}
\affiliation{Infrared Science Archive (IRSA), IPAC, 1200 E. California Blvd,
MS 314-6, California Institute of Technology, Pasadena, CA 91125 USA}
\author[0000-0002-1466-5236]{Moira Jardine}
\affiliation{School of Physics and Astronomy, University of St. Andrews, North
Haugh, St Andrews KY16922, UK}
\author[0000-0002-8863-7828]{Andrew Collier Cameron}
\affiliation{School of Physics and Astronomy, University of St. Andrews, North
Haugh, St Andrews KY16922, UK}
\author[0000-0002-3656-6706]{Ann Marie Cody}
\affiliation{NASA Ames Research Center, Space Sciences and
Astrobiology Division, MS245-3, Moffett Field, CA 94035 USA}
\author{Lynne A. Hillenbrand}
\affiliation{Astronomy Department,
California Institute of Technology, Pasadena, CA 91125 USA}
\author[0000-0002-5971-9242]{David Barrado}
\affiliation{Centro de Astrobiolog\'ia, Dpto. de
Astrof\'isica, INTA-CSIC, E-28692, ESAC Campus, Villanueva de
la Ca\~nada, Madrid, Spain}
\author[0000-0002-0493-1342]{Ethan Kruse}
\affiliation{NASA Goddard Space Flight Center, Greenbelt, MD 20771, USA}
\author[0000-0003-0501-2636]{Brian P. Powell}
\affiliation{NASA Goddard Space Flight Center, Greenbelt, MD 20771, USA}

\email{stauffer@ipac.caltech.edu}

\begin{abstract}

Using K2, we recently discovered a new type of periodic photometric
variability while analysing the light curves of members of Upper Sco
(Stauffer \etal\ 2017). The 23 exemplars of this new variability type
are all mid-M dwarfs, with short rotation periods. Their phased light
curves have one or more broad flux dips or multiple arcuate structures
which are not explicable by photospheric spots or eclipses by solid
bodies. Now, using TESS data, we have searched for this type of
variability in the other major sections of Sco-Cen, Upper
Centaurus-Lupus (UCL) and Lower Centaurus-Crux (LCC).  We identify 28
stars with the same light curve morphologies. We find no obvious
difference between the Upper Sco and the UCL/LCC representatives of
this class in terms of their light curve morphologies, periods or
variability amplitudes. The physical mechanism behind this
variability is unknown, but as a possible clue we show that the
rapidly rotating mid-M dwarfs in UCL/LCC have slightly different
colors from the slowly rotating M dwarfs - they either have a blue
excess (hot spots?) or a red excess (warm dust?).

One of the newly identified stars (TIC242407571) has a very striking
light curve morphology. At about every 0.05 in phase are features
that resemble icicles,  The ``icicles'' arise because there is a
second periodic system whose main feature is a broad flux dip. Using
a toy model, we show that the observed light curve morphology results
only if the ratio of the two periods and the flux dip width are
carefully arranged.

\end{abstract}

\section{Introduction}

Young stellar objects (YSOs) are often highly variable.   That
variability was in fact one of the defining characteristics of T~Tauri
stars (Joy 1945). However, almost all stars, even the Sun, are
variable if monitored with enough accuracy.   We are now in an era
when the whole sky is being  monitored on a regular basis from the
ground at good accuracy and cadence, and when much of the sky has been
monitored with very good accuracy and cadence by NASA's Transiting 
Exoplanet Survey Satellite (TESS; Ricker  \etal\ 2015). 
We are also in an era where ESA's Gaia satellite (Gaia Collaboration 2018) has
provided the ability to accurately identify all of the members of many
nearby young stellar associations where we had previously been limited
to only the high mass members or only very small samples of the whole
population.  These capabilities open up the possibility to discover
new types of variability in young stars, and perhaps thereby to learn
more about their formation process.

In a series of papers, we have recently used light curves from the K2
mission (Howell \etal\ 2014) to conduct an extensive survey of the 
rotational evolution of
low mass stars (Rebull \etal\ 2016a,b, 2017, 2018, 2020; Stauffer \etal\
2017, 2018a).  In the  process of conducting that survey, the two lead
authors (LMR and JRS) visually examined the K2 light curves of several
thousand young stars.   As a by-product of that effort, we discovered
a new class of photometrically variable, young stars (Stauffer \etal\
2017, 2018b; Rebull \etal\ 2016b, 2018). The majority of the stars so
identified came from our analysis of $\sim$1500 members of the $\sim$10
Myr old Upper Sco association.   

In this paper, we have examined light
curves from TESS of more than 3000 candidate members of the other two
major portions of Sco-Cen - the Upper Centaurus-Lupus (UCL) and
Lower-Centaurus-Crux (LCC) associations.  Both associations are
believed to be about $\sim$16 Myr old (Pecaut \& Mamajek 2016), thus allowing us to
determine if the variability characteristics we found in Upper Sco
change significantly as the stars move down their pre-main sequence (PMS) 
tracks.   

In Section 2, we describe in more detail the defining 
characteristics of the new
variability class.  The original sample of candidate UCL and LCC
members and published data we collected for those members is described
in Section 3.   The process we used to analyse the more than 3000 stars and
their light curves, and the sample of newly identified stars with
scallop-shell light curves or persistent flux dip light curves is
presented in Section 4.  Section 5 provides some physical  characterization of
the stars we have identified, and Section 6 compares some properties of
these stars to the previously identified scallop-shell and persistent flux dip
stars from Upper Sco.   In Section 7, we
describe a color-color diagram that could potentially help pinpoint
the physical mechanism(s) that result in the highly structured light
curves we see.  Finally, 
we discuss one particularly intriguing star
in Section 8 whose light curve appears to combine a scallop-shell
waveform at one short period, and a flux dip (or dips) not due to the
transit of a solid-body at another short period.  

\section{A Brief Discussion of Nomenclature}

The stars that are the subject of this paper exhibit a type of variability
unknown prior to 2017.  There is as yet no certain physical mechanism to
explain their variability (see, e.g., G\"unther \etal\ 2020 and references therein).  
In our discovery paper (Stauffer \etal\ 2017; 
hereafter S17), we
grouped the stars into three categories of light curves: ``scallop-shells,"
persistent flux dips, and transient flux dips.   Despite grouping the stars
into these categories, however, we suspect there is just one underlying
physical mechanism, with the three categories perhaps representing how
that mechanism manifests itself over a range of some (unknown) parameter.
The names were meant to capture the most characteristic morphological
feature of the phased light curve of each group:
\begin{itemize}
    \item scallop-shell -- the phased light curves of these stars show multiple 
    undulations,
    with no strong preference for upward of downward flux changes.  In some cases, the
 phased light curve has the appearance of the lip of a scallop shell.
 \item persistent flux dip -- the phased light curves of these stars show flux
dips whose shape and depth do not appear to vary significantly on day to
month timescales.  Often just one flux dip is present, but in some cases
several identifiable dips are present.   The dips are sometimes superposed
on light curves that appear to be that of normal spotted stars (often 
sinusoidal in shape), with the period associated with the dips equal to
that for the spotted-star waveform.
\item transient flux dip -- the phased light curves of these stars show flux
dips whose shape and depth vary significantly with time.   The depths can
vary on both day and month timescales.   These dips are usually superposed
on spotted-star light curves, with the dip period and spot period appearing
to be the same.
\end{itemize}

Zhan \etal\ (2019) identified a number of young stars with scallop-shell
light curves in early TESS data; most of these stars were members of young,
nearby moving groups.   Rather than use the scallop-shell short-hand to
describe these stars, they preferred a more fact-based nomenclature 
including ``highly structured,"  ``comprised of many Fourier components" and
``multi- and sharply-peaked."   We do not disagree with these descriptions,
and they are in some ways better.    However, we still prefer scallop-shell
for its brevity and we will use that term for the remainder of the paper.

Finally, we note that there is another class of variable PMS stars that
now commonly goes by the name ``dipper" (to a large extent equivalent with
AA Tau type variables -- Bouvier \etal\ 1999; Morales-Calderon \etal\ 2011; Cody
\etal\ 2014).   However, this category does not
overlap with our stars, and presumably has a very different physical mechanism
producing the variability.   Dippers instead are very young, PMS stars in almost
all cases with active accretion from a primordial disk and strong IR excesses;
their flux dips are generally believed to arise from dust in the primordial
disk itself or possibly from dust ``blobs" in transit from the disk towards
the star.  (Also see Ansdell \etal\ 2020.)

\section{Overall Sample Selection and Basic Data}

At least four groups have used the Gaia DR2 (Gaia Collaboration 2018) catalog 
to identify low mass members
of UCL and LCC (Zari \etal\ 2018; Damiani \etal\ 2019; Kounkel \& Covey 2019;
Goldman \etal\ 2018).  For high mass stars, Pecaut \& Mamajek (2016) provide a
well-documented set of probable UCL/LCC members.   For our analysis, 
we chose to merge the member lists from Zari \etal\ (2018),
Damiani \etal\ (2019), and Pecaut \& Mamajek (2016) to produce our catalog 
of UCL and LCC candidate.   The Zari \etal\ and Damiani 
\etal\ lists
have many stars in common, but also have a significant number of stars found only
in one list or the other.  We did not distinguish between those cases, but simply
adopted all of the stars as candidate members.

The Gaia DR2 catalog provides accurate coordinates for
all of the stars in our sample. Using those coordinates, we
downloaded all available near and mid-IR photometry for our stars from
the archives for 2MASS (Skrutskie \etal\ 2006), WISE/AllWISE (Wright \etal\ 2010), CatWISE 
(Eisenhardt \etal\ 2020), unWISE (Meisner \etal\ 2019), Spitzer
(Werner \etal\ 2004)
SEIP\footnote{http://irsa.ipac.caltech.edu/data/SPITZER/Enhanced/SEIP/overview.html},  
and AKARI (Murakami \etal\ 2007) . We obtained optical broadband
photometry from Gaia DR1 (Gaia Collaboration 2016) and DR2 (Gaia Collaboration 2018),  
PAN-STARRS1 for some stars (Chambers \etal\ 2016), APASS (Henden \& Munari 2014), 
NOMAD (Zacharias \etal\ 2005), 
the Southern Proper Motion Program (Girard \etal\ 2011), and
the GSC-II (Lasker \etal\ 2008). Pecaut \& Mamajek (2016) also
provide optical magnitudes.  
A spectral type was available in the literature for just one of our 
stars discussed here.  

We used all of the photometry to
produce spectral energy distributions (SEDs) for the stars in Tables 1 and
2.   As was true for the original S17 sample, based on their optical
colors and low extinctions,  all of the newly identified stars of interest
appear to be mid-to-late M dwarfs, and based on these SEDs,
all but a few appear not to have a detected IR excess.  The SEDs for
the new sample are provided in the Appendix.

\section{Newly Identified Stars of Interest}

We obtained ELEANOR (Feinstein \etal\ 2019) 30-minute cadence 
light curves for all of
the candidate members of UCL and LCC as described in the previous
section. As in our earlier papers, we selected the 
`best available' light curve version, this time from the 
products provided by ELEANOR.
We used the Lomb-Scargle (LS; Scargle 1982) approach as
implemented by the NASA Exoplanet Archive Periodogram
Service\footnote{http://exoplanetarchive.ipac.caltech.edu/cgi-bin/Periodogram/nph-simpleupload}
(Akeson \etal\ 2013). We also used the Infrared Science Archive (IRSA)
Time Series
Tool\footnote{http://irsa.ipac.caltech.edu/irsaviewer/timeseries},
which employs the same underlying code as the Exoplanet Archive service,
but allows for interactive period selection.

\begin{figure}[ht]
\epsscale{0.9}
\plotone{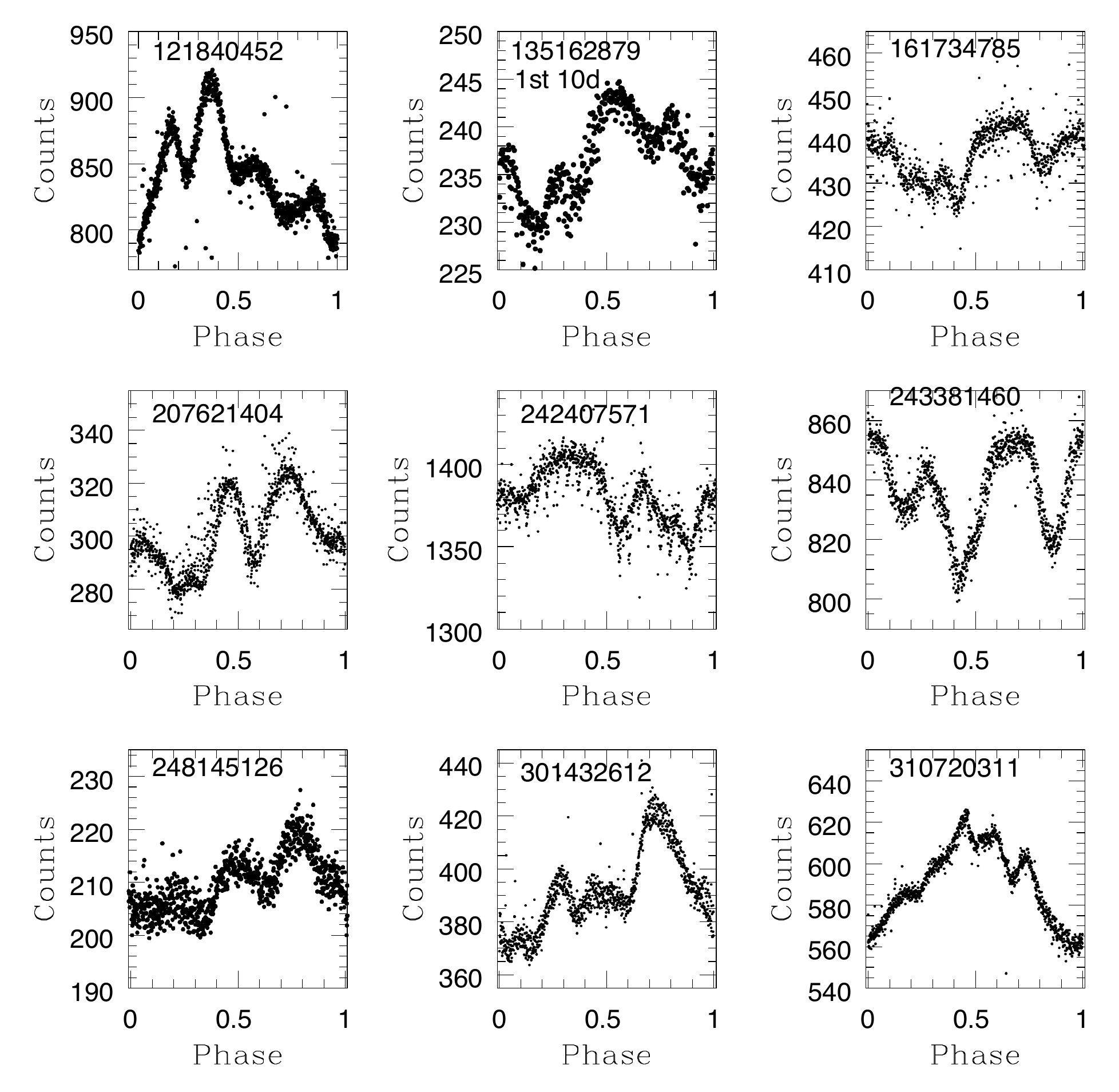}
\caption{The phased light curve  of the first nine stars in
Table 1 with scallop-shell phased TESS light curves.
\label{fig:superfigure1}}
\end{figure}

\begin{figure}[ht]
\epsscale{0.9}
\plotone{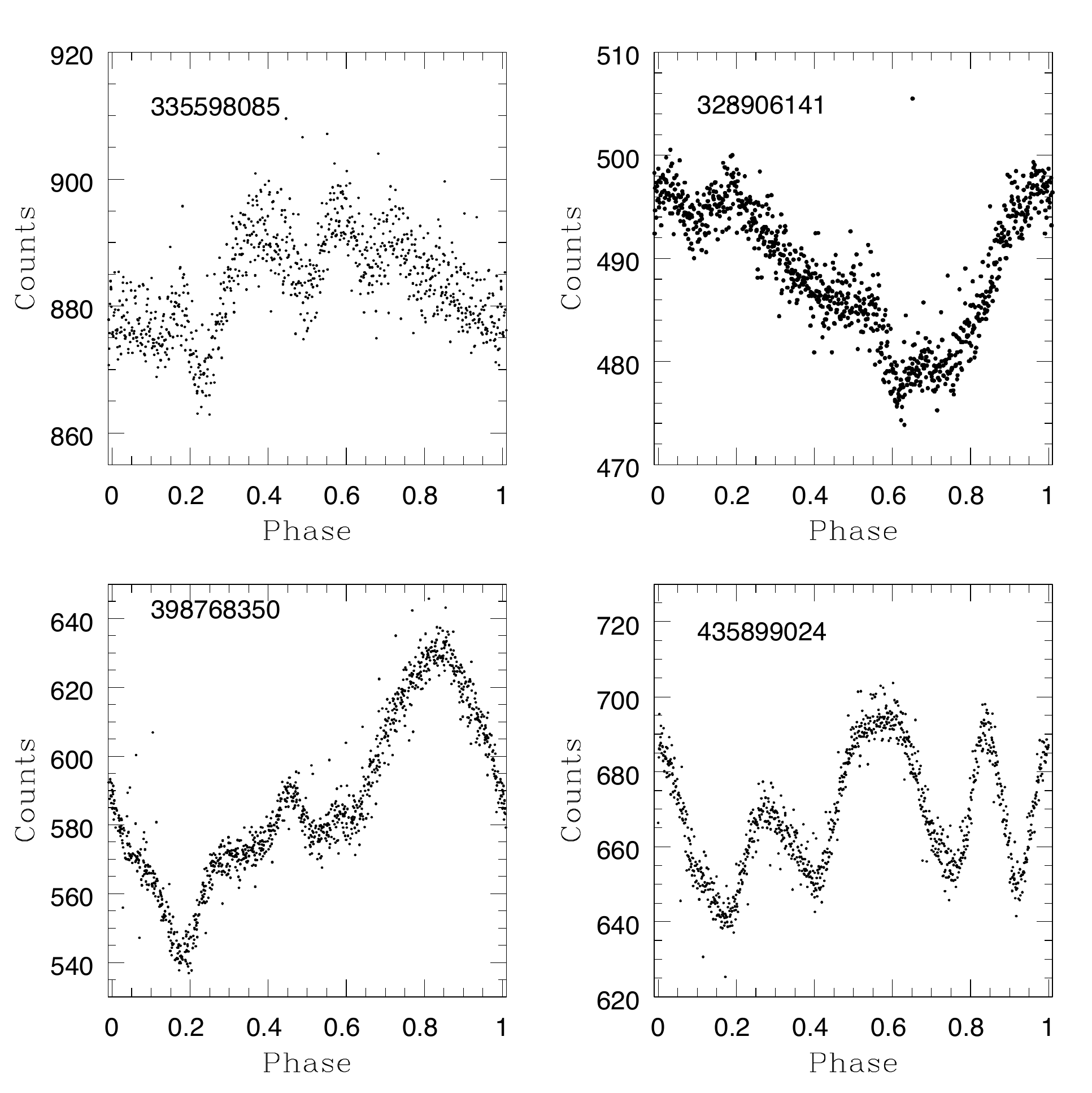}
\caption{The phased light curves of the remaining four stars in Table 1 with 
scallop shell morphologies.
\label{fig:superfigure2}}
\end{figure}

\begin{figure}[ht]
\epsscale{0.9}
\plotone{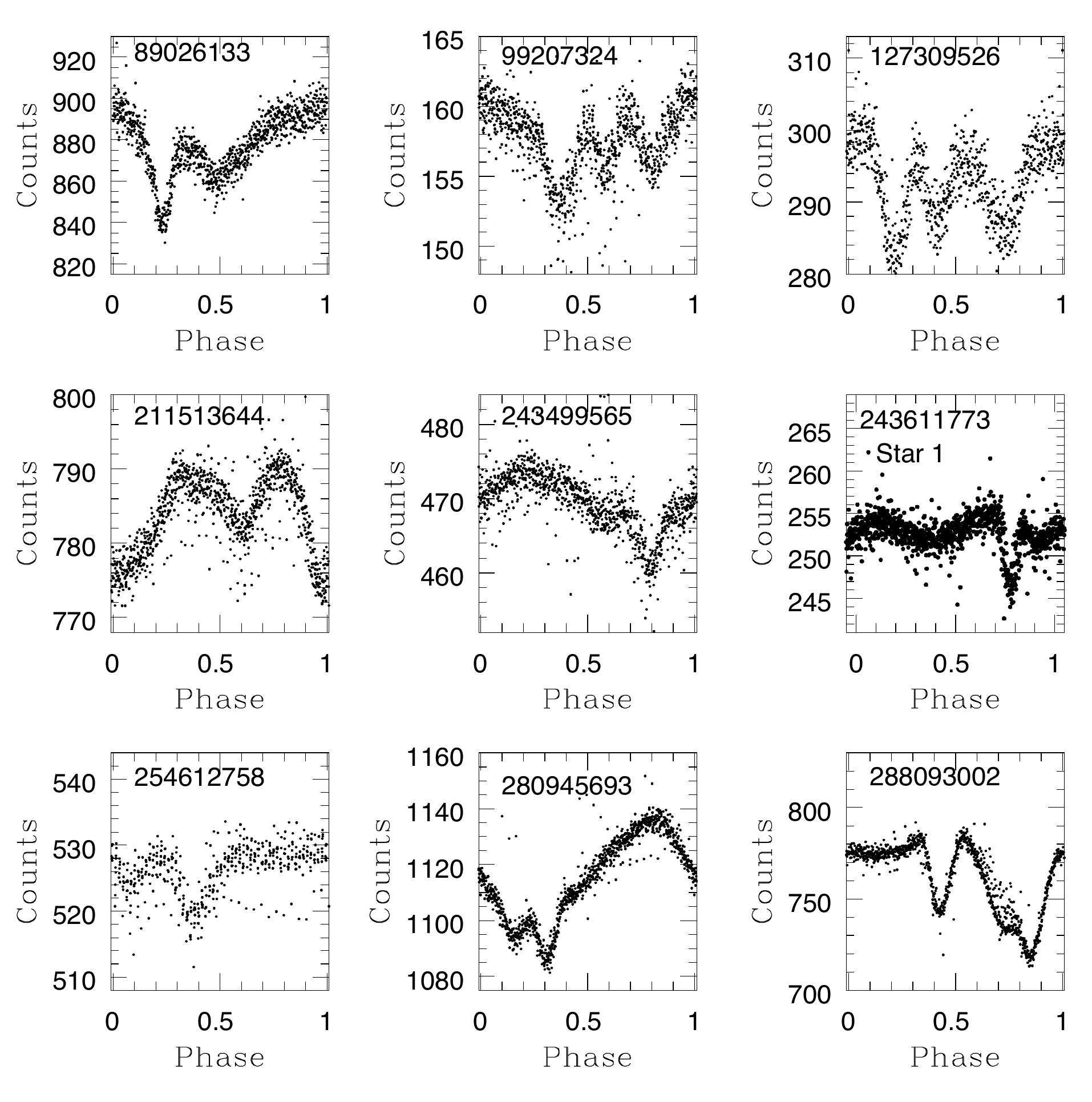}
\caption{The phased light curve of the first nine stars in Table 2 whose
TESS light curves show one or more persistent flux dips in their phased light
curves.
\label{fig:superfigure3}}
\end{figure}

\begin{figure}[ht]
\epsscale{0.9}
\plotone{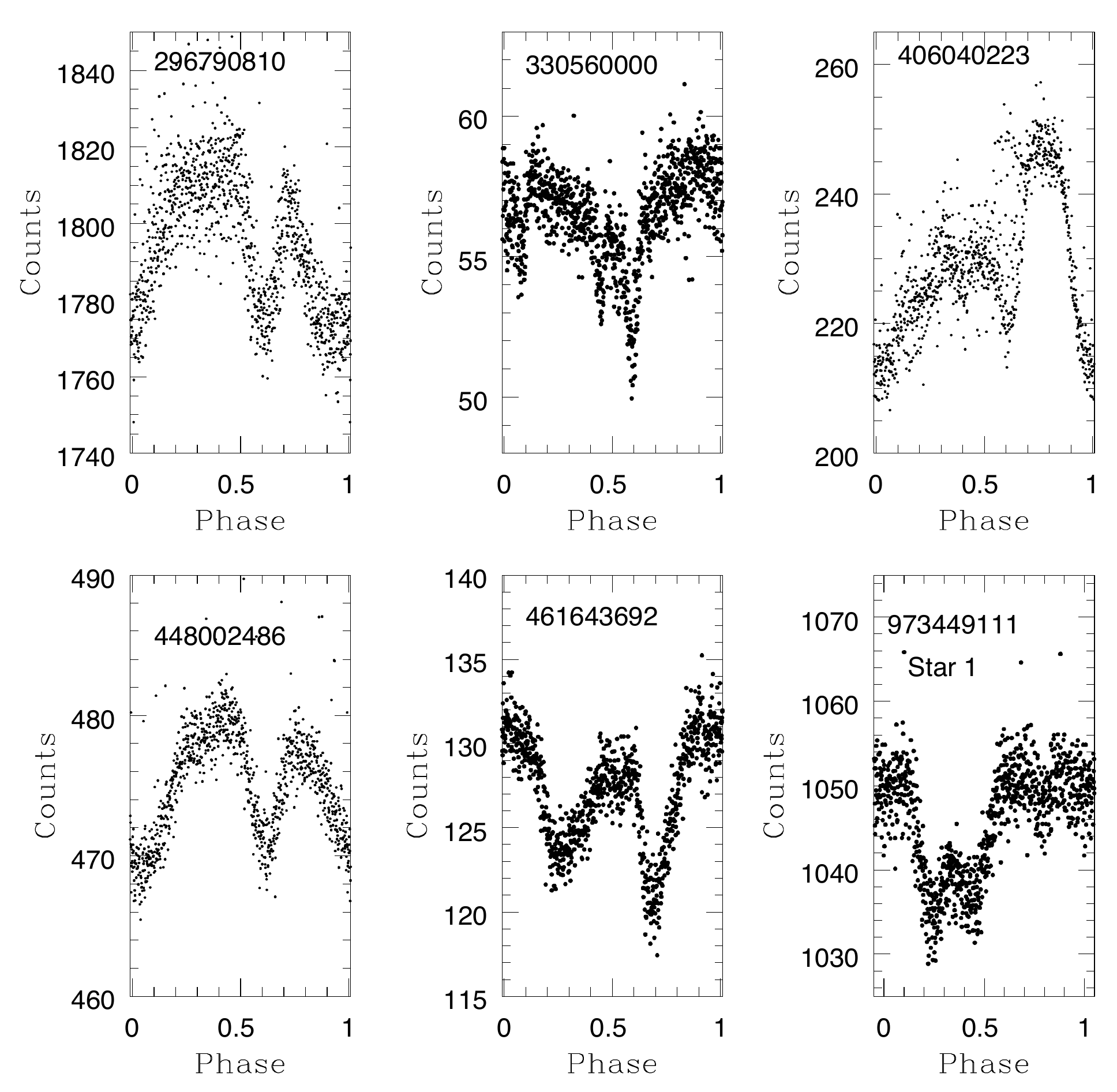}
\caption{The remaining six stars in Table 2 having
persistent, broad flux dips in their phased light curves.
\label{fig:superfigure4}}
\end{figure}

Two of us (LMR and JRS) then visually
examined all of the original light curves and the LS periodograms and
phased light-curves.   We flagged all stars whose light curves were 
unusual in some way (e.g., had characteristics that could put them into the
scallop-shell or flux-dip categories), and then conducted a more thorough analysis of
those stars; in some cases, this included downloading other light
curve versions from MAST, most often the CDIPS (Bouma \etal\ 2019)
light curve.    One of the outcomes of this effort was the
identification of a set of stars that we believe are good examples of
the scallop-shell and persistent flux-dip classes. Table 1 provides
the list of stars we place in the scallop-shell category; Table 2
provides the stars in the persistent flux dip category.  
In both tables, the 
light curve version for ELEANOR is included.  ELEANOR light curves 
can be PCA, principal 
component analysis; COR, corrected; or RAW. In a few cases, the
star could have been placed in either category, and we somewhat
arbitrarily chose what seemed the better classification\footnote{TIC 135162879 is a scallop
but maybe has an additional
variable flux dip at phase $\sim$0.3. TIC 99207324 and 211513644 are the most
ambiguous flux dip stars and could also have been placed in scallop shells, but the
dips are very triangular and there does seem to be a discernible 
out-of-dip portion of the light curve, landing them here instead.}.
The phased light curves of all of the stars in the two tables
are shown as Figures \ref{fig:superfigure1}-\ref{fig:superfigure4}.  The rotation periods of
the entire set of UCL and LCC candidate members and other analysis of
those stars will be reported in a separate paper (Rebull \etal\
in preparation).   

\floattable
\begin{deluxetable*}{lcccccccccc}
\tabletypesize{\footnotesize}
\tablecolumns{10}
\tablewidth{0pt}
\tablecaption{UCL/LCC Candidate Members with ``Scallop-Shell" Light Curves\label{tab:basicdata}}
\tablehead{
\colhead{TIC } &
\colhead{RA } &
\colhead{Dec} &
\colhead{Gaia $G$} &
\colhead{Gaia $B_P-R_P$} &
\colhead{Gaia Parallax} &
\colhead{P}  &
\colhead{Amplitude\tablenotemark{a}} &
\colhead{TIC contam\tablenotemark{b}} &
\colhead{LC version\tablenotemark{c}} & \colhead{Notes}\\
& \colhead{(deg)}& \colhead{(deg)}
& \colhead{(mag)}& \colhead{(mag)}
& \colhead{(milliarcsec)}
& \colhead{(days)} }
\startdata
121840452 & 226.3205 & -38.2367 & 14.029 & 2.992 & 8.499 & 0.3787 & 14.9\% & 0.26 & COR & P2 = 0.205 \\
135162879 & 185.5549 & -41.8007 & 14.986 & 3.001 & 7.668 & 0.3776 &  7.2\% & 0.005& COR & \\
161734785 & 190.4468 & -51.1685 & 15.179 & 2.895 & 6.767 & 0.3487 &  4.6\% & 0.27& COR &  \\
207621404 & 206.8866 & -55.5505 & 14.597 & 2.817 & 6.995 & 0.4516 &  15.\% & \ldots & PCA &\\
242407571 & 213.6035 & -45.9454 & 13.082 & 2.737 & 8.952 & 0.4721 &  \ldots & 0.51 & COR &\\ 
243381460 & 205.0069 & -43.8159 & 13.349 & 2.677 & 7.736 & 0.3684 &  5.4\% &0.02& PCA & \\
248145126 & 193.9389 & -44.8643 & 15.223 & 3.076 & 7.790 & 0.3504 & 11.1\% &0.09 & PCA & \\
301432612 & 178.5196 & -58.0447 & 13.870 & 2.732 & 9.246 & 0.5042 &  8.4\% &0.32 & COR & \\
310720311 & 185.4716 & -63.7926 & 13.084 & 2.565 & 9.379 & 0.6649 &  9.7\% &0.42 & RAW &  \\
328906141 & 211.4775 & -52.4334 & 15.036 & 2.948 & 6.732 & 0.4629 &  4.1\% &0.18& PCA & \\
335598085 & 194.8984 & -68.1337 & 13.086 & 2.858 & 9.412 & 0.6607 &  3.1\% &0.38& PCA & \\
398768350 & 180.3328 & -56.8174 & 13.333 & 2.638 & 9.225 & 0.4899 &  15.\% &\ldots & PCA & \\
435899024 & 193.0948 & -64.3109 & 14.057 & 3.014 & 9.699 & 0.36335 & 7.9\% &1.04 & RAW &\\
\enddata
\tablenotetext{a}{Amplitude of variability of the phased light curve;
given as a percent =  [(max counts - min counts)/average counts]x100.}
\tablenotetext{b}{Value of contamination ratio from TESS Input Catalog (TIC), e.g., 
ratio of flux from nearby objects that falls in the aperture of the target 
star, divided by the target star flux in the aperture; see Stassun \etal\ (2018). }
\tablenotetext{c}{ELEANOR provides several different light curve versions. The one we used is listed in this column.}
\end{deluxetable*}
\noindent

\begin{deluxetable*}{lccccccccccc}
\tabletypesize{\footnotesize}
\tablecolumns{10}
\tablewidth{0pt}
\tablecaption{UCL/LCC Candidate Members Whose Light Curves Show Persistent, Broad Flux Dips\label{tab:basicdata2}}
\tablehead{
\colhead{TIC } &
\colhead{RA } &
\colhead{Dec} &
\colhead{Gaia G} &
\colhead{Gaia B-R} &
\colhead{Gaia Parallax} &
\colhead{P}  &
\colhead{Amplitude\tablenotemark{a}} &
\colhead{TIC contam\tablenotemark{b}} &
\colhead{FWZI\tablenotemark{c}} & \colhead{LC version\tablenotemark{d}}  \\
 & \colhead{(deg)}& \colhead{(deg)}
& \colhead{(mag)}& \colhead{(mag)}
& \colhead{(milliarcsec)}
& \colhead{(days)}   }
\startdata
 89026133 & 230.6437 & -35.0705 & 13.477 & 2.820 & 7.157 & 0.4665 & 4.5\% & \ldots& 0.21 & RAW  \\
 99207324 & 236.5251 & -35.4197 & 15.615 & 3.112 & 9.664 & 0.7926 & 3.5\% & 0.11 & 0.22 & COR \\ 
127309526 & 216.9984 & -43.4414 & 14.449 & 2.965 & 6.001 & 0.4341 & 9.0\% &\ldots& 0.29 & RAW \\
211513644 & 217.0398 & -49.2627 & 13.904 & 2.703 & 7.856 & 0.6344 & 1.9\% &0.41& ... & PCA \\
243499565 & 206.1492 & -47.1038 & 14.504 & 2.698 & 7.684 & 1.1942 & 1.9\% & 0.03& 0.14 & COR \\
243611773 & 207.0679 & -44.0440 & 15.041 & 3.214 & 6.740 & 0.3778 & ...  & 0.06& ... & COR \\
254612758 & 236.2847 & -44.4355 & 13.826 & 2.708 & 5.754 & 0.5933 & 3.0\% & 0.48& 0.28 & RAW  \\
280945693 & 174.0702 & -69.4643 & 13.474 & 2.972 & 10.145 & 0.6363 & 2.3\% & \ldots & 0.14 & PCA \\
288093002 & 185.1167 & -54.5937 & 13.883 & 2.767 & 9.569 & 0.43145 & 7.3\% & 0.47& 0.20 & COR \\
296790810 & 176.6070 & -66.6932 & 13.756 & 3.089 & 9.306 & 0.3713 & 2.4\% & \ldots& 0.23 & PCA \\
330560000 & 213.6021 & -51.0554 & 16.188 & 3.236 & 7.857 & 1.3899 & 11\% & 0.19& 0.09 & PCA \\
406040223 & 193.8415 & -58.7782 & 14.967 & 3.184 & 9.034 & 0.3241 & 10.4\% & 0.20& ... & COR \\
448002486 & 185.4086 & -69.1439 & 14.741 & 3.188 & 9.038 & 0.3594 & 2.1\% & 0.17& 0.28 & PCA \\
461643692 & 224.5963 & -33.7376 & 15.878 & 3.148 & 6.125 & 0.4113 & 11.4\% & 0.06& 0.26 & COR \\
973449111 & 199.8996 & -62.5769 & 13.921 & 2.647 & 9.397 & 0.6210 & 3.1\% & 1.41& ... & PCA \\
\enddata
\tablenotetext{a}{Amplitude of variability of the phased light curve;
given as a percent =  (max counts - min counts)/(max counts)x100. }
\tablenotetext{b}{Value of contamination ratio from TESS Input Catalog (TIC), e.g., 
ratio of flux from nearby objects that falls in the aperture of the target 
star, divided by the target star flux in the aperture; see Stassun \etal\ (2018). }
\tablenotetext{c}{Full-width zero intensity (FWZI) of the most prominent flux dip. (See Stauffer
\etal\ 2017 for more discussion of FWZI in the context of scallop shells.)}
\tablenotetext{d}{ELEANOR provides several different light curve versions. The one we used is listed in this column.}
\end{deluxetable*}

\clearpage

\section{Physical Properties of the Stars in Tables 1 and 2}

Unlike the stars identified as having scallop-shell light curves or
having persistent flux dips in the K2 Upper Sco sample, the stars so
identified in UCL and LCC have very little characterization in the literature.
Only one of them has a published spectral type.  Most have no mention
in the literature other than having been identified as candidate
members of UCL and LCC based on Gaia DR2 data.  We can now provide
some characterization  of their properties, however, based primarily
on their DR2 photometry and astrometry.

\begin{figure}[ht]
\epsscale{0.9}
\plotone{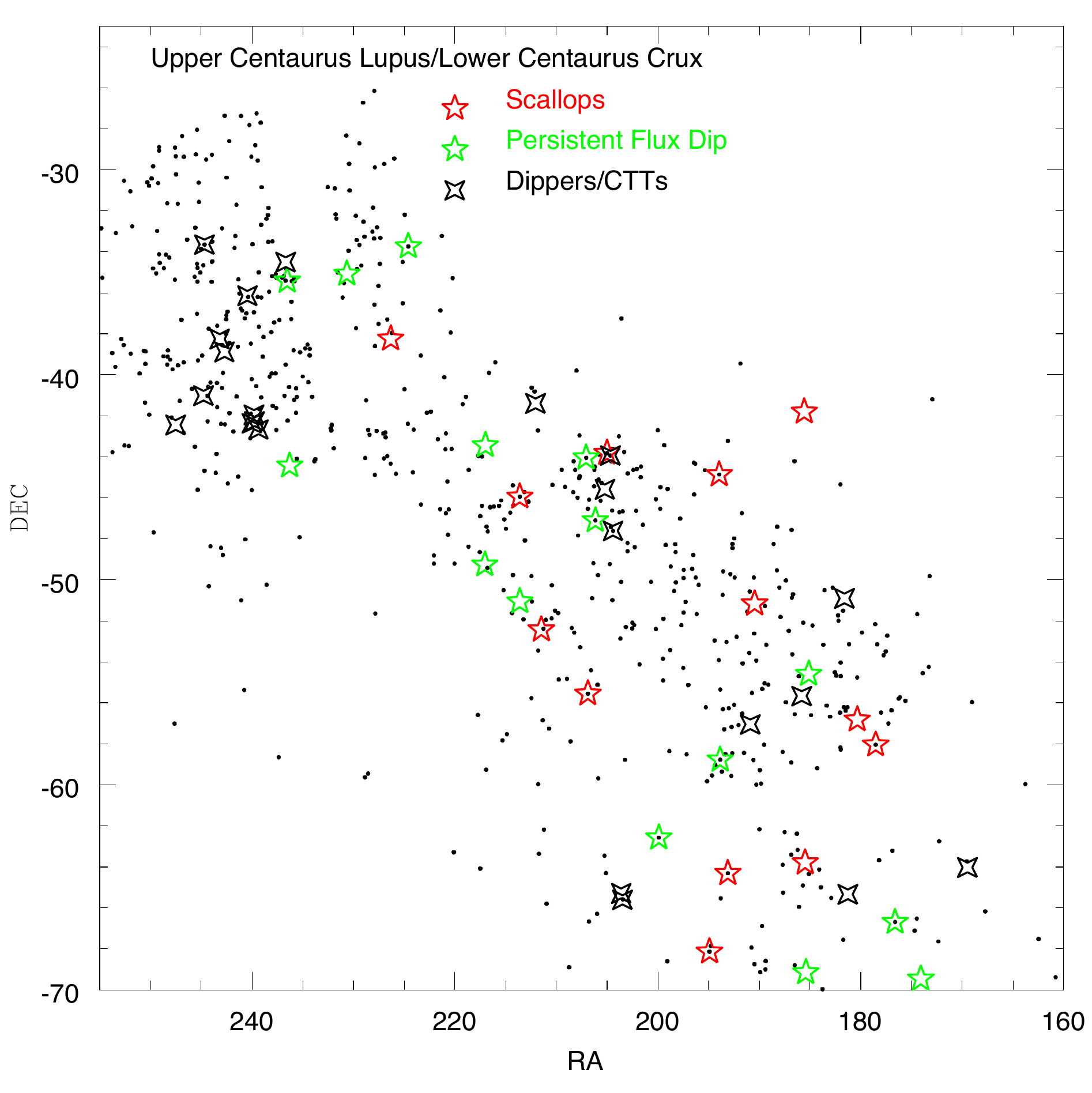}
\caption{Spatial plot, showing the location on the sky of the candidate UCL/LCC members with
TESS periods from our preliminary analysis of the ensemble.  
The stars in Tables 1 and 2 are plotted as five point star symbols.
Additionally, stars identified as having CTT light curves (accretion bursts, 
AA Tau-like flux dips, or similar) are shown as four-point star symbols.   The portion of
UCL with RA $> 235\arcdeg$  includes a significant population of CTTs but only two of our stars with
highly structured light curves (the division between UCL and LCC is roughly
at RA = 210$\arcdeg$, with UCL to the left in the plot and LCC to the right). This
could indicate that the stars in that portion of UCL are, on average, younger than
the stars in the other half of UCL or the stars in LCC.
\label{fig:spatial}}
\end{figure}

Figure \ref{fig:spatial} shows the location on the sky of the stars in Tables 1 and
2.  Also marked in this diagram are all of the UCL and LCC candidate
members with TESS periods based on our preliminary analysis of the ELEANOR
light curves (the complete, final set of periods will be provided in
Rebull \etal\ in preparation); we also 
indicate the location of stars
whose light curves suggest they are actively accreting classical T~Tauri 
(CTT) stars 
(``dippers", bursters, etc.)  based on our preliminary analysis of
the full dataset.
The stars of Tables 1 and 2 are
scattered throughout the region, but seem to be more prevalent in LCC
than UCL.   On the other hand, the stars with CTT light curves appear
to be somewhat more prevalent in UCL.   The higher rate of CTTs
could be interpreted as indicative
that, on average, UCL is somewhat younger than LCC.

\begin{figure}[ht]
\epsscale{0.9}
\plotone{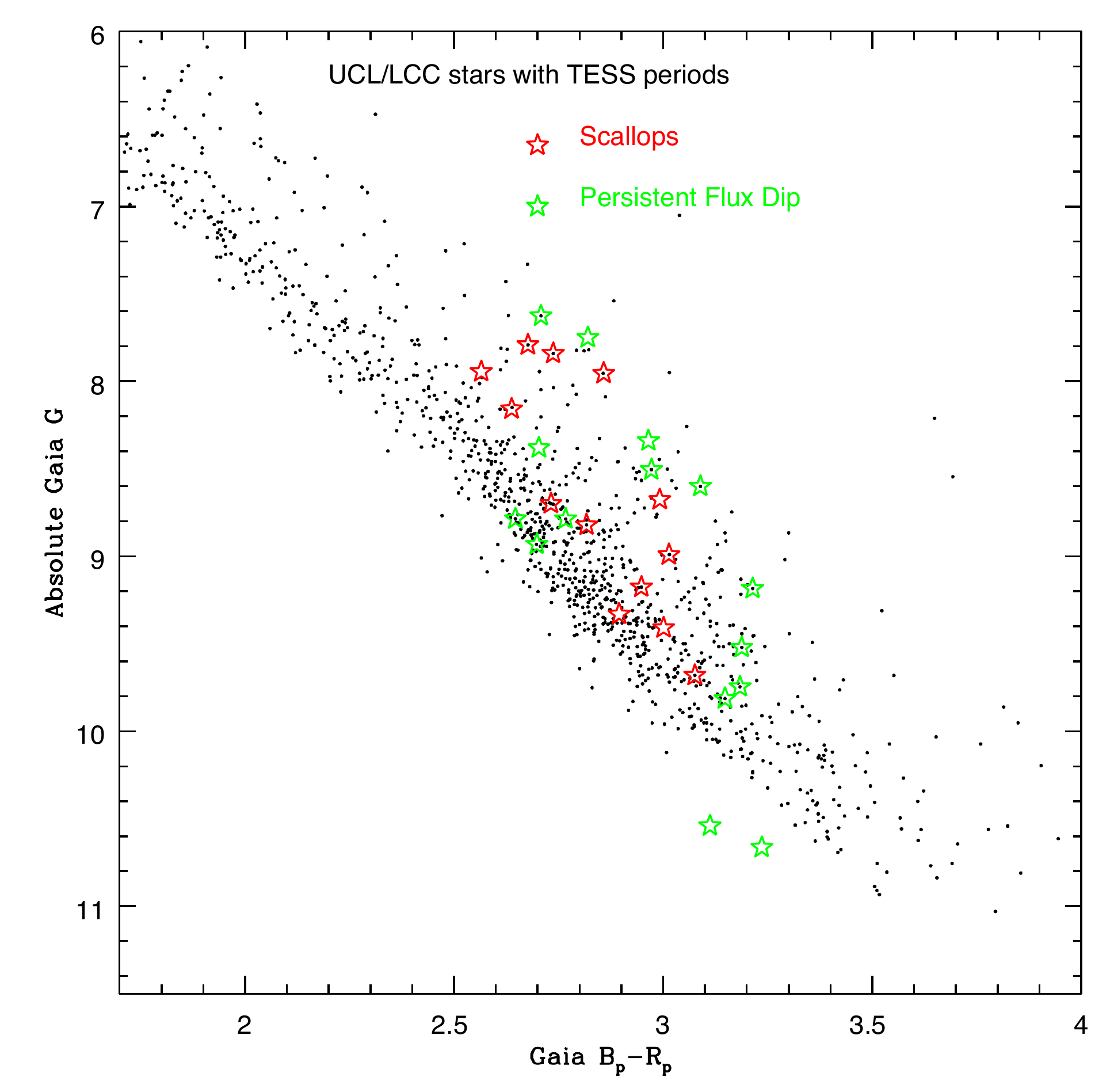}
\caption{Absolute $G$ magnitude versus Gaia $B_p-R_p$ color for UCL/LCC candidate 
members from our preliminary analysis.
The stars in Tables 1 and 2 are again plotted as star symbols.  Many of the
scallop-shell and persistent flux dip stars are well-displaced above the 
single-star locus in the diagram,
either because they are younger or that they are members of binary systems.
\label{fig:gaiacmd}}
\end{figure}

Figure \ref{fig:gaiacmd} shows a Gaia-based color-magnitude 
diagram (CMD) for our UCL/LCC sample, again 
highlighting the stars of Tables 1 and 2 among our preliminary analysis of 
the rest of the UCL/LCC stars.   Based on their $B_p-R_p$
colors and  the calibration of the PARSEC 16 Myr isochrone (Chen \etal\ 2014), 
TESS light
curves are available for stars down to nearly 0.1 \msun. However, for targets
redder than about $B_p-R_p$ = 3.2, in most cases, the S/N of the light
curves is too low to positively identify a star as having a scallop
shell or persistent flux-dip light curve morphology.   The $B_p-R_p$ colors of the
stars we include in Tables 1 and 2 are consistent with their having
spectral types of about M2.5 to M5.   The CMD appears to show
both a fairly well-defined single-star locus and a significant number
of stars displaced well above that locus; the stars displaced above
the single-star locus could be binaries or they could be younger than
16 Myr.  More than half of the scallop shell and persistent flux dip stars 
are located amongst the set of
stars displaced above the single-star locus.   More than half of the stars with 
CTT light curves are also displaced well above the single-star locus in the CMD;
in their case, youth is the most likely explanation.

\begin{figure}[ht]
\epsscale{0.9}
\plotone{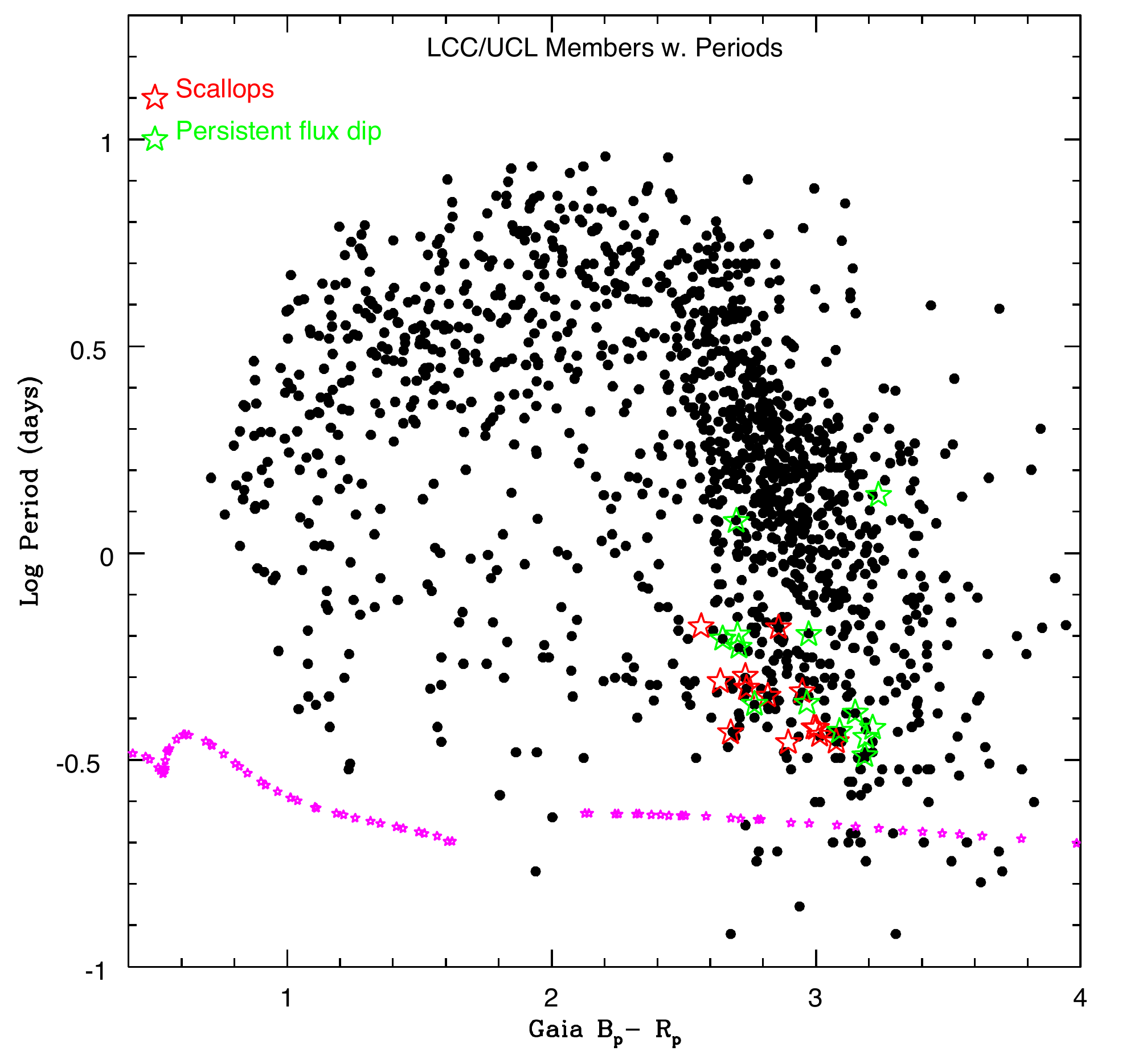}
\caption{Log of the rotation period versus Gaia $B_p-R_p$ color for UCL/LCC candidate
members from our preliminary analysis.   
The stars in Tables 1 and 2 are plotted as star symbols.   The dashed
magenta line shows an estimate of the rotational breakup rate, using the formula
from Bouvier (2013) and the PARSEC 16 Myr isochrone.  The scallop-shell and
persistent flux dip stars are all rapid
rotators, but their rotation periods are generally two or more times longer than the breakup
rate.
\label{fig:periodgaiacolor}}
\end{figure}

Figure \ref{fig:periodgaiacolor} shows the same set of stars as in the previous two figures,
but this time in a rotation versus color plot (again with the preliminary analysis
of the ensemble of UCL/LCC stars).  The dashed magenta
line is the breakup period, calculated using a formula from Bouvier
(2013) with the PARSEC 
16 Myr isochrone as input.  This diagram will
be discussed in much more detail in  Rebull \etal\ (in preparation).   For the
purposes of the present paper, the important points are that the scallops and
persistent flux dip stars are only found amongst the mid to late M
dwarfs, and that they concentrate at the rapidly rotating end of the
distribution.  However, while they are rapidly rotating, they are -- on
average -- not particularly close to the breakup period.   The rotation
periods for the scallop shell stars (0.273 d $< P <$ 0.605 d) are
generally somewhat shorter than those for the persistent flux dip
stars (0.324 d $< P <$ 1.39 d), but there is considerable overlap
between the two distributions.

We have measured the amplitude of variability for our stars, and we
provide those numbers in Tables 1 and 2.  For the scallops, amplitudes
range from 2.2\% to 15\%, whereas the persistent flux-dip stars from Table 2 
have amplitudes of 1.9\% to
11\%.  The amplitudes we measure are probably lower 
limits to the true 
amplitude, as a result of dilution of star light by nearby stars. 
Stassun \etal\ (2018) provide estimates of contamination, which are also 
included in the tables above. Where this ratio is available, it is typically small
for these stars. Taking all of this into account, while we believe the 
general trends shown by
the measured amplitudes, the amplitudes for individual stars 
may be large than we report. 
 
If one places a tight box in Figure \ref{fig:periodcolordm} around the stars of Table 1 and 2 (2.5 $< B_p-R_p <$ 3.2
and $-0.5 < \log P < -0.1$), one can extract a set of rapidly rotating dM stars from our
preliminary analysis possibly
useful for comparison to the scallops and persistent flux dip stars.    Doing so, and
excluding our stars of interest, we find 110 such stars.   
The frequency for which 
a rapidly rotating dM stars shows the scallop/flux dip characteristics is then
28/138, or about 20\%.   We use this set of rapid rotators more in \S 8.

\section{Comparison to the Scallop Shell Stars in Upper Sco}

Despite being monitored by different satellites, 
the similarities between the scallop shell and persistent 
flux-dip stars in UCL/LCC, studied here with TESS, 
and those presented earlier in Upper Sco, based on K2 data, are substantial. 
For both datasets,
the intrinsic faintness and distance of the stars of interest put those stars
near the flux limit for TESS or K2 to provide useful light curves. 
Because UCL/LCC is older than Upper Sco, the stars of interest are intrinsically
fainter; for LCC, that is compensated for by the closer mean distance to Earth, but
UCL is about the same mean distance as Upper Sco and so our ability to identify
scallop shell and flux-dip stars is most compromised there.

One possibly significant difference between UCL/LCC and Upper Sco is the ratio of scallop-shell
to persistent flux-dip stars identified; in Upper Sco, that ratio is
15/8 or 1.8($^{+0.57}_{-0.34}$).  In UCL/LCC,
the ratio is 13/15 or 0.86($^{+0.05}_{-0.13}$).   
While small number statistics is a problem, this
could suggest that the lifetime of scallop shells is significantly shorter than
that for persistent flux-dip stars.   The fact that a few persistent flux-dip stars are still present
at Pleiades age (but no scallop shells with age $>$ 50 Myr are known) is consistent
with this (Rebull \etal\ 2016b).  

The period distributions in UCL/LCC for our stars with highly structured light curves are only slightly
different from their cousins in Upper Sco.   For Upper Sco, periods for scallops 
range from 0.26 day $<  P  <$ 0.78 day,
with a median of 0.46 day; for the persistent flux dip stars, the period range is
0.48 day $<  P  <$ 1.54 day, with a median of 0.62 day.  The same quantities for
UCL/LCC (given in the previous section) are very similar.   That may, in fact,
be a little surprising given that low mass stars generally spin up on their path to the
main sequence.    

While we do not have spectral types in UCL/LCC, we do have accurate Gaia colors
and we know there is little reddening 
(see, e.g., Pecaut \& Mamajek 2016, where LCC and UCL have similar 
reddening based on 144 and 195 stars, respectively; dropping three 
anomalously high $A_{V}$ values from UCL, then the mean for both is $A_{V}\sim0.15\pm$0.20).   
The Gaia $B_p-R_p$ colors for the scallops
and persistent flux-dip stars imply spectral types ranging from M2.5 to M5.  The measured
spectral types in Upper Sco for scallops and persistent flux dip stars
range from M3 to M6.3, with all but a couple being
M5 or earlier.   The red end of the spectral type range in both cases is
almost certainly set by the photon count rates at K2 and TESS.

The amplitudes of variability in Upper Sco show the same wide range as in
UCL/LCC, except that the maximum amplitude found in our sample was 21\%,
compared to the 15\% in UCL/LCC.   We find no significant difference in the
flux dip depths or widths between the two groups. 

\clearpage

\section{Photometric Properties of the Stars in Tables 1 and 2}

As described earlier, the lack of published data (e.g., spectra, high
resolution imaging, etc) for our UCL/LCC stars limits our ability to
attempt to use their properties to constrain possible theoretical models
to explain their variability.   However, UCL/LCC does have one very 
important positive attribute relative to Upper Sco.  The low mass stars in 
Upper Sco have both relatively large reddening and large star-to-star
reddening differences (see, e.g., discussion in Rebull \etal\ 2018 and 
references therein).   The variable reddening adds considerable scatter
to any diagram where color is used for one axis.  UCL/LCC, on the other
hand, has quite low reddening ($A_V \sim$ 0.15 mag).   With the availability of
the very accurate Gaia optical photometry and the good 2MASS and WISE
near- and mid-IR photometry, it may be possible to identify color differences
associated with our stars having highly structured light curves.  

\begin{figure}[ht]
\epsscale{0.9}
\plotone{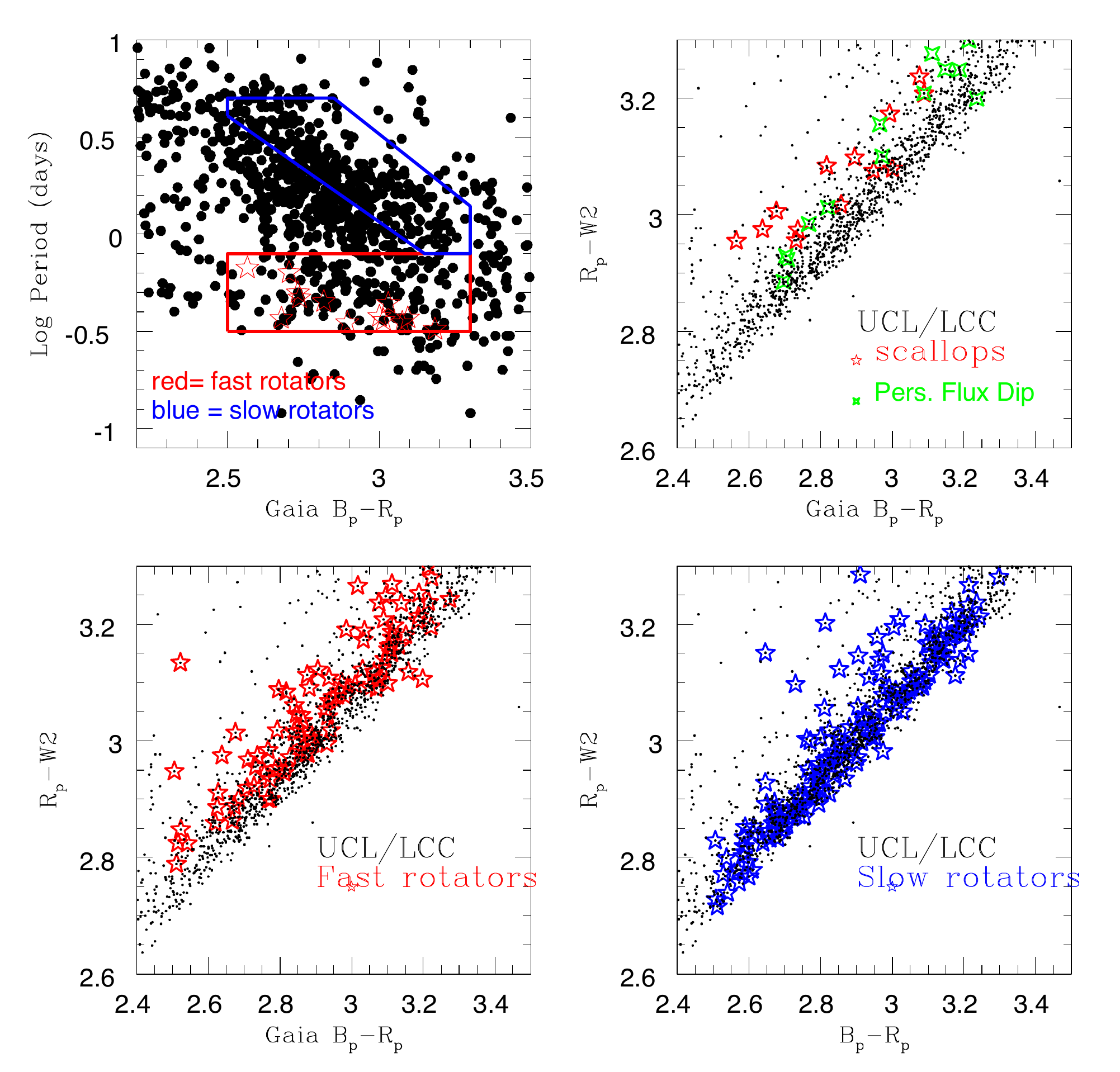}
\caption{ (a) log Period vs.\ color plot for dM members of UCL/LCC with periods;
(b) $B_p-R_p$ vs.\ $R_p-[4.6]$ plot for the stars in panel (a), showing that the scallop
shell light curves are systematically displaced from the bulk of the dMs;
(c) similar plot for the rapid rotators; (d) similar plot for the slow
rotators.  
\label{fig:periodcolordm}}
\end{figure}

We have indeed found that there is a shared photometric anomaly among
the rapidly rotating dM stars in UCL/LCC which may provide a clue to
the photometric variability of the stars in Tables 1 and 2.   
Our evidence for this is provided in Figure \ref{fig:periodcolordm}.   In order to investigate
whether the rapid rotators in general (or the scallop shell stars in
particular) have unusual photometric properties, we used a plot of
period versus color to define several sets of stars, which we illustrate
in the top-left panel of Figure \ref{fig:periodcolordm}.  The stars with scallop-shell and
persistent flux-dip morphology all have colors of 2.5 $< B_p-R_p <$ 3.3 and log $P$
between $-$0.1 and $-$0.5.  All of the black dots in the plot represent the
full sample of stars with periods from our preliminary analysis.  
The region outlined in blue defines the
set of more slowly rotating stars; the region outlined in red defines the
rapid rotators (from which we have removed all the stars in Tables 1 and 2).
The top-right panel of Figure \ref{fig:periodcolordm} compares the location in a plot of
$B_p-R_p$ vs.\ $R_p-[4.6]$ (where [4.6] is the WISE-2 magnitude) for the 
scallops and persistent flux-dip stars versus the entire
sample.   Our stars of interest are clearly displaced either blueward in
$B_p-R_p$ or redward in $R_p-[4.6]$ (or both) relative to the full sample.  The
bottom half of Figure \ref{fig:periodcolordm} shows similar plots for the slowly rotating and
fast rotating stars.   The rapid rotators appear to be displaced from the
main locus in a similar amount as for the scallops and persistent flux-dip stars; the
slow rotators track the main locus of stars.   The bottom line, therefore,
is that all of the rapidly rotating, mid-to-late dM members have colors
that are significantly different from slow rotators.  

How can we interpret the color anomaly of the rapid rotators?  If the
displacement is primarily due to their having bluer $B_p-R_p$ colors, this
could arise from their having a significant contribution to their light
from hot spots (or plages).   If the displacement is primarily due to their
having redder $R_p-[4.6]$ colors, that could arise from their being a small
amount of warm dust in the system.   Unfortunately, without additional
data, we cannot differentiate between those possibilities at this time.

\section{TIC 242407571: There is Another Period}

\begin{figure}[ht]
\epsscale{0.9}
\plotone{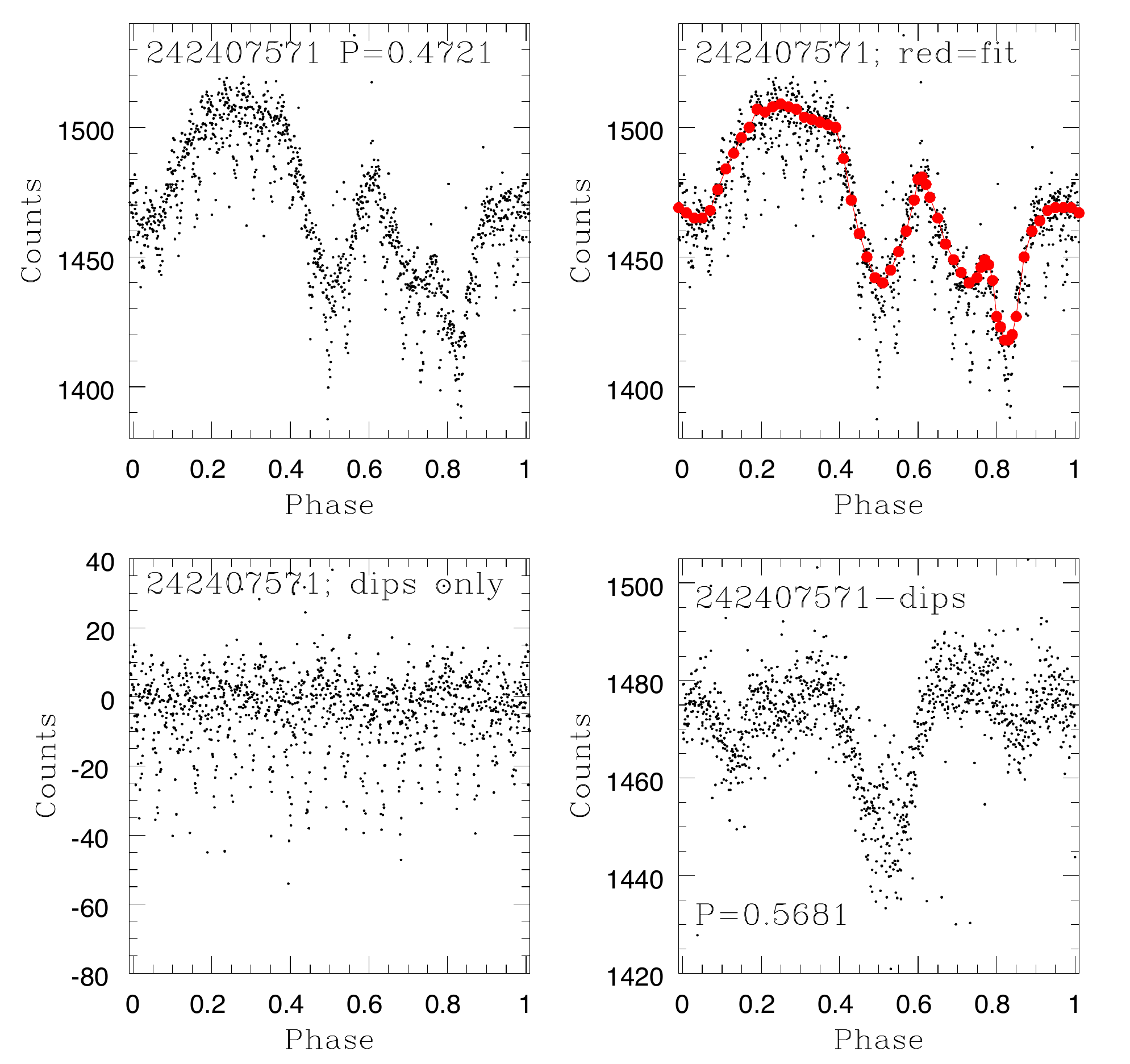}
\caption{Various versions of the TIC 242407571 light curve, as 
described in the text.
\label{fig:tic242407571}}
\end{figure}

One of the most striking light curve morphologies shown in Figures 1-4 is
the phased light curve for TIC 242407571.   We reproduce that light curve
here in Figure \ref{fig:tic242407571}a.   The most unusual and perplexing 
features of Fig. \ref{fig:tic242407571}a are the
icicle-like features ``hanging" down from the scallop-shell waveform; the
``icicles'' appear to be both very narrow and reasonably equally spaced in phase.
However, it is extremely unlikely that they are part of the scallop-shell
waveform.   None of the other stars in Tables 1 or 2 show
features like this, nor, for that matter, for all of the Upper 
Sco (Rebull \etal\ 2018) stars we have discussed, nor in the
rest of the stars in 
UCL and LCC (Rebull \etal\ in preparation), at least as part of 
our preliminary analysis.  
While we do not have a spectrum for TIC 242407571, its
Gaia and 2MASS colors indicate that it should have a spectral type of
about M4.   Using the PARSEC 2.3 isochrone for a 16 Myr population, the
Gaia color of $B_p-R_p$ = 2.737 converts to a mass of 0.40 \msun, and an estimated
radius of 0.80 \msun.  It is located well above the single star locus in the CMD
shown in Figure \ref{fig:periodcolordm}, compatible with it being a nearly equal mass binary.

In order to investigate the apparent icicles more closely, we have first made
a fit to the scallop shell waveform, shown in Figure \ref{fig:tic242407571}b.   We then 
subtracted this fit from the original phased waveform, yielding the 
pure icicle light curve shown in Figure \ref{fig:tic242407571}c.   We retained the original
MJD timestamps for each point during this process, and so we next 
ran a period search on this light curve using a box-least squares
algorithm (also available at the NASA Exoplanet Archive Periodogram
Service).   A strong period of $P$ = 0.5681 day was
found; Figure \ref{fig:tic242407571}d shows the icicle light curve phased to
this period.   The phased light curve shows a relatively broad, apparently
stable (in shape and depth) flux dip with a depth of about 2.5\% and full-width
zero intensity (FWZI) 
of about 0.22 in phase, plus two weaker outrigger flux dips separated
by about 120 degrees in phase (on either side) from the primary dip.

How does the $P$=0.5681 day flux dip waveform result in the evenly
spaced, nearly equal length icicle curtain
shown in Figure \ref{fig:tic242407571}a?   
There seems to be two primary drivers: a) the period of the flux
dips and the period of the scallop-shell waveform have a ratio very close to 1.20
and b) the width of the main flux dip is wide enough
that $\sim$4 TESS data samples are affected by the dip each period.   
For every five dip periods, there are six scallop-shell periods, and
then the pattern repeats every $\sim$2.8 days.  Because
the ratio of the two periods (1.203) is very close to 6/5, the flux dips from the 0.568 day period
align when phased to the 0.4721 day scallop-shell period, at least
for the limited $\sim$20 day duration of each TESS campaign.  If any
one of the three parameters (TESS sampling frequency; ratio of the two periods;
width of the flux dip) were significantly different, the evenly spaced,
comb-tooth appearance of the icicles would not have happened.  

\begin{figure}[ht]
\epsscale{0.9}
\plotone{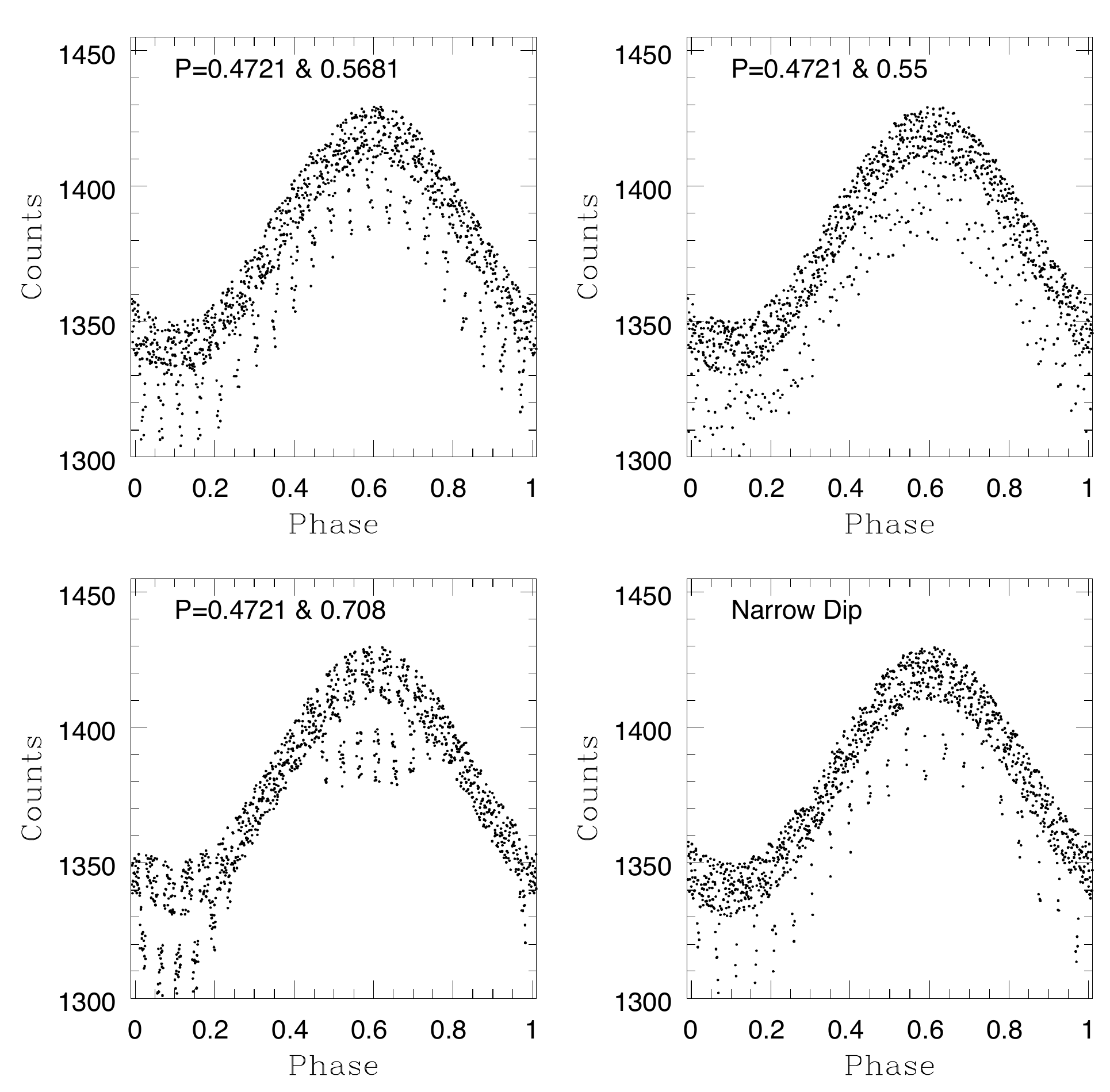}
\caption{Various versions of the model light curves, as 
described in the text.
\label{fig:models}}
\end{figure}

We have created
a simple model to illustrate this point, where we have replaced the scallop-shell
waveform with a sine wave, and we have then added a flux dip at a different period
from the light curve, retaining the exact observing cadence as for the real light
curve.   Figure \ref{fig:models} shows four instances of this model.   In the top-left panel,
a model with the same second period as the real star and a flux dip similar to the
real flux dip was used; the resultant phased light curve shows icicles very similar
to that observed.  The top right panel shows the same model, except using a second
period of 0.55 days, resulting in a random-appearing set of data points below
the sine wave.  The bottom left panel instead uses a period of 0.708 days,
corresponding to 1.5 times the sine wave period of 0.4721d.  We again see icicles,
but of varying lengths and with a large gap between groups of them (the result
of the dip width no longer being appropriate for the period ratio).   Finally,
the lower right panel shows what happens when the flux dip is made narrower; this
results in icicles that are less well sampled.    

Two questions immediately  come to mind at this point.   First, is the
near alignment of the two periods by chance, or driven by
some presumably gravitational process?   Second, what is the nature of
the material causing the flux dips which result in the 0.5681 day
periodicity?

While having the two periods so well aligned (so that their ratio is
near 1.20) would seem to suggest that a physical process was involved,
there could be another interpretation.   If the two periods were not
so aligned, the 0.568 day flux dips would just result in a randomly positioned
set of anomalously low points when phased to the 0.4721 day period (see
Figure \ref{fig:models}c).  Our
automated software did not pick up the 0.5681 day period because it provides
too small a signal in the LS periodogram as compared to the signal from
the scallop-shell waveform.  Because we had to scan the plots for thousands
of stars, it is possible our visual scanning of the phased light curves
might have attributed the haze of low lying points as noise or data
artifacts.   Thus, the period alignment may have been an integral part
of identifying this star as interesting -- and other similar cases (but
lacking an alignment to produce an ordered appearance of the low-lying
points) may be present in other Sco-Cen light curves if analysed thoroughly.

The dip characteristics for TIC 242407571 are entirely
compatible with
what we have measured for the persistent flux
dip stars in Table 2.  Therefore, whatever physics is responsible for the
persistent flux dip stars may well explain the $P$ = 0.5681 flux dips in
TIC 242407571.

The model for TIC242407571 that has the fewest unusual assumptions is that
it is in fact a binary system.   One star exhibits the scallop shell light
curve, and that star has a rotation period of 0.4721.   The second star, which
is probably a relatively wide binary companion, exhibits the $P$ = 0.5681 day persistent
flux dip morphology. TIC242407571 is displaced well above the single star locus
in Figure \ref{fig:periodcolordm}, in support of it being a binary.  
This model does not require any significant interaction
between the two stars, and assumes that the fact that the ratio of their
two periods of $\sim$1.20 is by chance. However, this star does not have 
exceptionally high astrometric excess noise nor exceptionally high 
re-normalised unit weight error.

\clearpage

\section{Summary}

By providing high quality, most-of-the-sky, deep, synoptic data
for millions of stars, TESS is an ideal tool for discovering, or characterizing,
new classes of photometric variability.  
We had previously used data from K2 to discover that
some or perhaps all rapidly rotating mid-dM stars have highly structured 
light curves that could not be explained by any of the existing physical
mechanisms inducing variability.    In this paper, we have used TESS data
to survey the light curves for thousands of low mass members of the 
UCL and LCC associations.   We found nearly 30
of those stars are striking new examples of the light curve class we identified
with K2.  In Figures \ref{fig:superfigure1}-\ref{fig:superfigure4} 
we show phased-folded light curves for these
stars, and in Figures \ref{fig:SEDS1}-\ref{fig:SEDS2} we provide plots of their SEDs.

The $B_p-R_p$ colors of the 28 UCL/LCC stars indicate that they
have spectral types of M2.5 to M5.   Their rotation rates range from 0.23 to
0.66 days; while quite rapid, except for two of the stars, those periods are
more than a factor of two longer than the breakup period.   Their amplitudes
of variability range from a few percent to 15\%; in particular, four of the
scallop-shell stars have amplitudes of $\sim$15\%. Such large amplitudes of variability will be
a challenge for any of the physical mechanisms to explain.  

The properties we have measured for this set of stars differ only slightly from
those we had previously measured for the discovery set of objects in the
Upper Sco association.   The one most significant difference, perhaps, is the
ratio of scallop shell to persistent flux dip stars -- in Upper Sco, that ratio
is 15/8, whereas in UCL/LCC the ratio is 13/15.   One explanation would be that
some stars have their light curve morphology evolve over time from more scallop-like
to more flux-dip like.   Another possibility is that the scallop-shell morphology
has a shorter lifetime than the persistent flux dip morphology.

In \S 8, we have shown that the rapidly rotating low mass stars
in UCL/LCC have colors that are either blue in $B_p-R_p$ or red in $R_p-[4.6]$ compared
to the slowly rotating UCL/LCC members.  The scallop-shell and persistent flux
dip stars share those anomalous colors; their colors do not seem to stand
out amongst the rapid rotators.   Blue $B_p-R_p$ colors could arise if hot spots
contribute significantly to the optical light from these stars.   Red $R_p-[4.6]$
colors would most naturally arise if these stars had a small amount of warm
dust.   

What is most needed at this point is a viable physical mechanism that can
recreate the variability properties we, and others, have measured.  But observers
can still help.   Red-optical spectra to measure radial velocities and the H$\alpha$\
emission profile at several phase points would be helpful.  Multi-color, synoptic photometry
of a few of the largest amplitude stars could help constrain dust grain properties.
A survey for additional examples of this light curve class for a well-defined 30-40
Myr cluster or moving group would help determine the lifetimes of the scallops and
persistent flux dip morphologies.

\begin{acknowledgements}

Some of the data presented in this paper were obtained from the
Mikulski Archive for Space Telescopes (MAST). Support for MAST for
non-HST data is provided by the NASA Office of Space Science via grant
NNX09AF08G and by other grants and contracts. 
This research has
made use of the NASA/IPAC Infrared Science Archive (IRSA), which is
operated by the Jet Propulsion Laboratory, California Institute of
Technology, under contract with the National Aeronautics and Space
Administration. This research has made use of NASA's Astrophysics Data
System (ADS) Abstract Service, and of the SIMBAD database, operated at
CDS, Strasbourg, France. This research has made use of data products
from the Two Micron All-Sky Survey (2MASS), which is a joint project
of the University of Massachusetts and the Infrared Processing and
Analysis Center, funded by the National Aeronautics and Space
Administration and the National Science Foundation. The 2MASS data are
served by the NASA/IPAC Infrared Science Archive, which is operated by
the Jet Propulsion Laboratory, California Institute of Technology,
under contract with the National Aeronautics and Space Administration.
This publication makes use of data products from the Wide-field
Infrared Survey Explorer, which is a joint project of the University
of California, Los Angeles, and the Jet Propulsion
Laboratory/California Institute of Technology, funded by the National
Aeronautics and Space Administration.
\end{acknowledgements}

\facility{TESS} \facility{Exoplanet Archive} \facility{IRSA}
\facility{2MASS} \facility{WISE}

\clearpage

\appendix

\section{Colors and Spectral Types}

The correspondence between $B_p-R_p$ and spectral type that we are using is derived 
from the data in Table~\ref{tab:bprpspty}, where the spectral types come from the 
literature. All of the 
stars discussed here have $B_p-R_p$ 2.5 to 3.25, and are therefore taken to be M2.5 to M5.

\begin{deluxetable*}{lcccccccc}
\tabletypesize{\footnotesize}
\tablecolumns{3}
\tablewidth{0pt}
\tablecaption{Correspondence between $B_p-R_p$ and spectral type \label{tab:bprpspty}}
\tablehead{
\colhead{Object } &
\colhead{Spectral Type } &
\colhead{$B_p-R_p$}  
}
\startdata
HII 2940 &  M0    &1.75\\
HII 624  &  M2    &2.23\\
HII 2601 &  M2    &2.205\\
HII 2602 &  M2.5  &2.435\\
HII 906  &  M3    &2.387\\
MT 61    &  M3    &2.889\\
HCG 456  &  M4    &2.888\\
VA208    &  M4.6  &3.299\\
VA203    &  M4.6  &3.150\\
VA 362   &  M5    &3.16\\
GJ905    &  M5    &3.531\\
HHJ 6    & $\sim$M6    &3.512\\
VB8      &  M7    &4.7542\\
\enddata
\end{deluxetable*}

\section{Spectral Energy Distributions}

\begin{figure}[ht]
\epsscale{1.}
\plotone{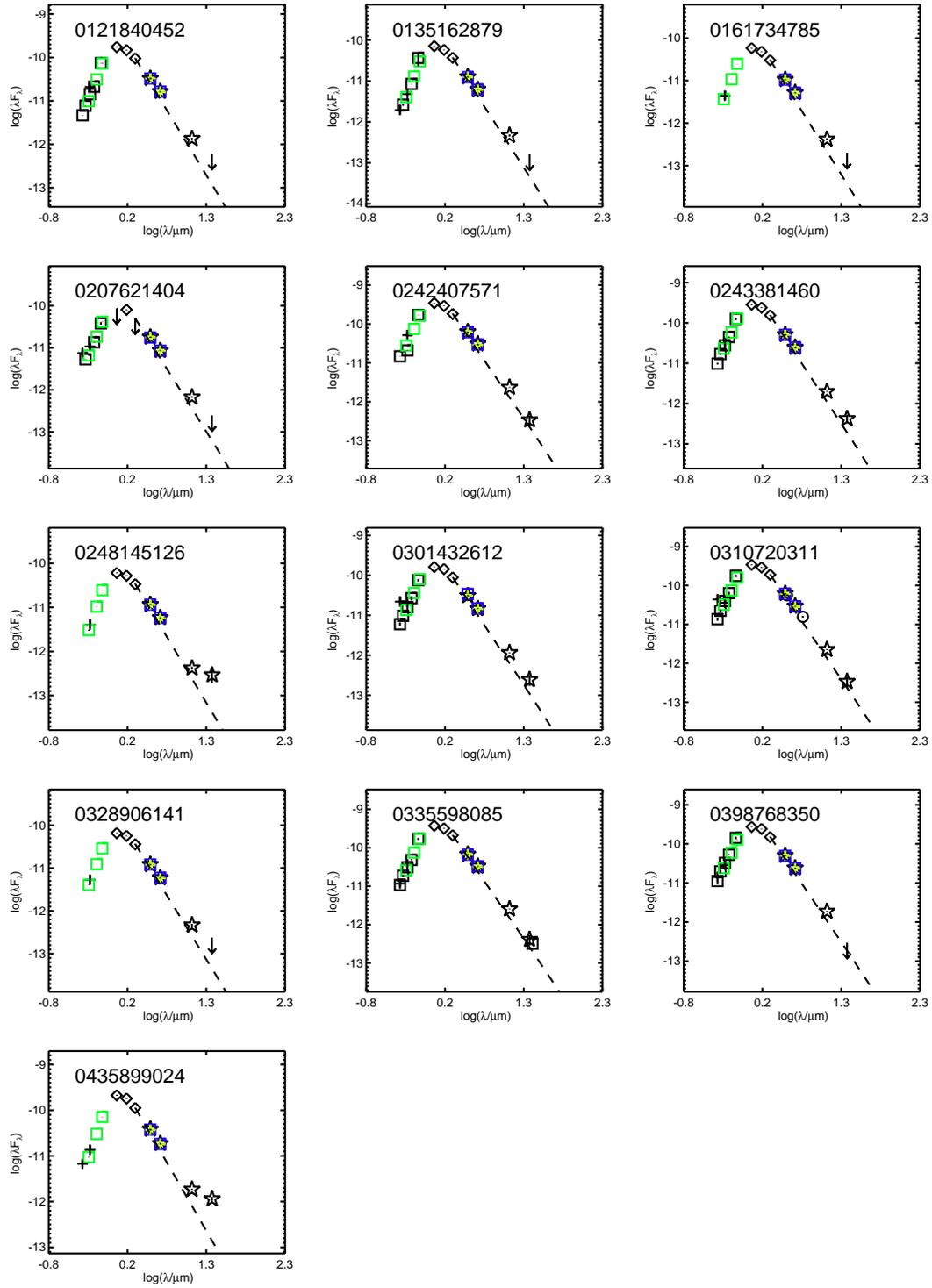}
\caption{Spectral energy distributions for the stars in Table 1.
See text for a description of the symbols.
\label{fig:SEDS1}}
\end{figure}

\begin{figure}[ht]
\epsscale{1.}
\plotone{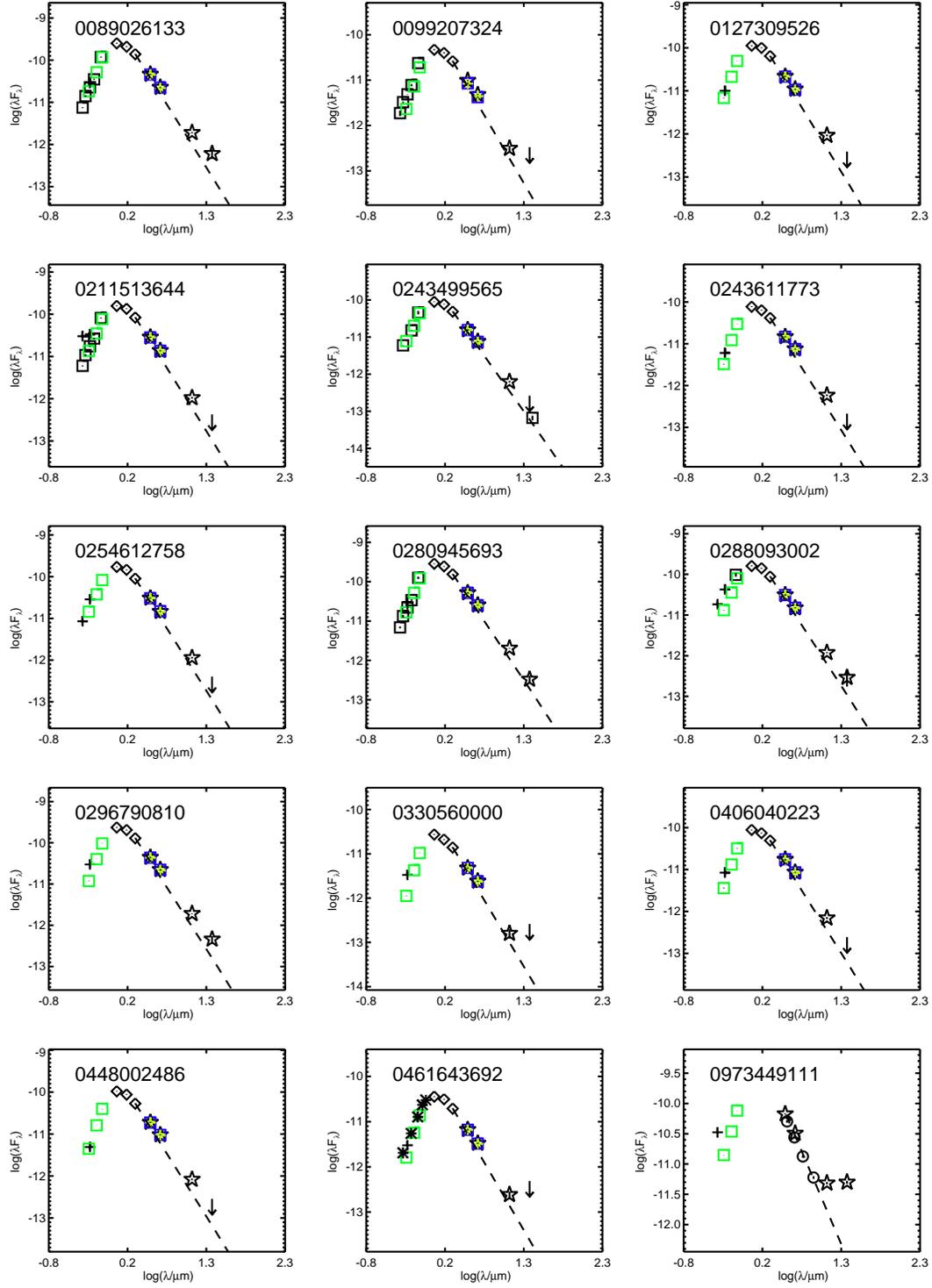}
\caption{Spectral energy distributions for the stars in Table 2.
\label{fig:SEDS2}}
\end{figure}

Spectral energy distributions for all the stars from Tables 1 and 2 are
provided here as Figure \ref{fig:SEDS1} and Figure \ref{fig:SEDS2}.
Plots are log $\lambda F_{\lambda}$ in cgs units (ergs s$^{-1}$
cm$^{-2}$) against log $\lambda$ in microns. Symbols: 
black square at short bands: APASS;  green square at short bands: Gaia DR2;
black asterisk at short bands: Pan-STARRS1; 
$+$: optical other literature (NOMAD, SPM, etc. not already plotted);
diamond: 2MASS $JHK_s$; circle: IRAC; black square at long bands: MIPS; 
stars: WISE from AllWISE; blue square in the mid-IR: CatWISE; 
green $+$ in the mid-IR: unWISE; 
arrows: limits; vertical bars (often smaller than the symbol) denote
uncertainties.  A line with a Rayleigh-Jeans slope is also shown as
the dashed line, extended from the observations at $K_s$ when available, 
or WISE-1. Note that this
is not a robust fit, but just to ``guide the eye.'' 
Almost all the stars have SEDs consistent with pure
photospheres; none have large IR excesses.

\end{document}